\documentclass{emulateapj}

\shorttitle{The Thermal Structure of $\gamma\,$Cas's Disk}
\shortauthors{T.\ A.\ A.\ Sigut \& C.\ E.\ Jones}

\begin{document}

\title{The Thermal Structure of the Circumstellar Disk Surrounding the
Classical Be Star $\gamma\,$Cassiopeia}

\author{T.\ A.\ A.\ Sigut \& C.\ E.\ Jones\\
Department of Physics and Astronomy, The University 
of Western Ontario, London, Ontario, N6A 3K7, Canada 
\email{asigut@astro.uwo.ca} \email{cjones@astro.uwo.ca}}

\slugcomment{Accepted for Publication by {\it The Astrophysical Journal}}

\begin{abstract}

%
%
We have computed radiative equilibrium models for the gas in the
circumstellar envelope surrounding the hot, classical Be star
$\gamma\,$Cassiopeia. This calculation is performed using a code
that incorporates a number of improvements over previous treatments
of the disk's thermal structure by \citet{mil98} and \citet{jon04};
most importantly, heating and cooling rates are computed with atomic
models for H, He, CNO, Mg, Si, Ca, \& Fe and their relevant ions. Thus,
for the first time, the thermal structure of a Be disk is computed for
a gas with a solar chemical composition as opposed to assuming a pure
hydrogen envelope.  We compare the predicted average disk temperature,
the total energy loss in H$\alpha$, and the near-IR excess with
observations and find that all can be accounted for by a disk that is
in vertical hydrostatic equilibrium with a density in the equatorial
plane of $\rho(R)\approx 3$ to $5\cdot\,10^{-11} (R/R_*)^{-2.5}\, \rm
g\,cm^{-3}$.  We also discuss the changes in the disk's thermal structure
that result from the additional heating and cooling processes available
to a gas with a solar chemical composition over those available to a
pure hydrogen plasma.

\end{abstract}


\section{Introduction}

Classical Be stars are non-supergiant B~stars that possess circumstellar
material in the form of an equatorial disk. While the circumstellar
disk is almost certainly a decretion disk of material from the star's
atmosphere, the detailed mechanism that creates and maintains such a
disk remains unclear \citep{por03,owo04}.  Rapid rotation of the central
B~star seems to play an important role, but there is still considerable
debate as to the extent \citep{tow04,fre05}.  Historically, the observational
evidence for such circumstellar material has been either spectroscopic
or polarimetric in nature, and the accepted observational definition of a
Be star has been the appearance (at the current or previous epoch)
of emission in the hydrogen Balmer lines. It has been recognized since
the days of \citet{str31} that recombination in a flattened disk could
reproduce the range of spectroscopically observed H$\alpha$ profiles.
In addition, the net (continuum) linear polarization observed in Be
stars is well explained by electron scattering from non-spherically
distributed circumstellar gas \citep{coy69,wat92}.

Beginning with the resolution of $\phi\,$Persei (B2~Vpe) at radio
wavelengths with the Very Large Array by \citet{dou92}, interferometry
has increasingly been used to spatially resolve circumstellar material. Be
star disks have been resolved at radio, near-IR, and optical wavelengths,
with these observations conclusively revealing the disk \citep{qui93,
qui97, tyc05, tyc06}.  The observations are consistent with the
suggestion of \citet{poe78} that Be star disks are geometrically quite
thin with opening angles of only a few degrees.

Currently, optical interferometric observations require theoretical
models of the emitting region in order to interpret the observed
visibilities. Often observers fit simple models with free parameters to
the data to describe the disk emissivity. However, this simple procedure
can be considerably improved by using a detailed model for the
thermal structure of the Be star disk. Such models naturally predict
the emissivity and opacity of the gas required to produce theoretical
spectroscopic images \citep{mil98,jon04,car05}.

$\gamma\,$Cassiopeia (HD~5394; B0~IVe) is a interesting classical Be star
which has a dense, cool, equatorial disk \citep{mil98}. This disk has been
resolved with optical interferometry \citep{ste95,tyc05}.  $\gamma\,$Cas
is likely the primary in a binary system \citep{har00,mir02} and has
unique X-ray characteristics \citep{smi04}.  Although $\gamma\,$Cas
is often quoted as ``the prototypical" Be star, it has become clear
that it possesses some unique characteristics.  Nevertheless, it is
a well-studied star making it an appropriate choice to test new codes
and compare results with previously published work.  In this current
paper, we extend the radiative-equilibrium models of \citet{mil98} and
\citet{jon04} for the early Be star $\gamma\,$Cas to a disk models with
a solar chemical composition.

\section{Calculations}

\subsection{Overview}

The calculations in this paper were performed with a new code, {\sc
bedisk}, which is loosely based upon the calculational approach of
\citet{mil98} and \citet{jon04}.  Models for the circumstellar material
were constructed assuming a density distribution falling as an $R^{-n}$
power-law in the equatorial plane, following the models of \citet{wat86},
\citet{cot87}, and \citet{wat87}.  By comparing to observations of the
infrared excesses of Be stars, these authors found power-law density
exponents in the range $n\approx\,2.0$ to $3.5$.  Perpendicular to the
equatorial plane, it was assumed that the gas is in vertical, isothermal,
hydrostatic equilibrium.  Given the disk density distribution and the
photoionizing radiation field from the central star, the equations of
statistical equilibrium were solved for the ionization state and level
populations of H, He, CNO, Mg, Si, Ca, and Fe subject to the constraints
of charge and particle conservation. Radiative transfer was handled
via the escape probability approximation, and it was assumed that the
dominant escape route for photons was perpendicular to the exponentially
stratified disk.  From this solution, the rates of energy gain and
energy loss at each computational grid point were obtained, and the local
temperatures were iteratively adjusted to enforce radiative equilibrium.

The main features and assumptions of this code are now discussed in detail.

\subsection{The {\sc bedisk} Code}

\label{sec:detail}

The circumstellar disk was assumed to be axisymmetric about the
star's rotation axis and symmetric about the mid-plane of the disk. In
all that follows, $R$ is used for the radial distance from the star's
rotation axis and $Z$, for the perpendicular height above the
equatorial plane. Thus the cylindrical co-ordinates of any disk location
are $(R,Z)$. If $R_*$ denotes the stellar radius, then the calculation domain
is $R_* \le R_i \le R_{\max}$ with $i=1\ldots\,n_r$, and $0 \le Z_j \le
Z_{\max}(R_i)$ with $j=1\ldots\,n_z$. Typically, $n_r$ is set to 60 with
$R_{\max}/R_*=50$, and $n_z$ is set to 40.

The code accepts a user-defined set of atomic models which list
the energy levels and the bound and free radiative and collisional
transitions for each atom and ion to be included in the calculation.
The set of atoms and ions and the total number of atomic levels and
radiative transitions used for the current calculations are listed in
Table~\ref{tab:atomic_models}. For this initial work, the number of
energy levels included for each atom and ion is similar to the list of
energy levels given by \citet{mor68}. Sources for the required atomic
data are given in Appendix~A. The abundances assumed for the various
elements can be found in Table~\ref{tab:abun} and are taken from the
accepted solar abundances of \citet{and89} and \citet{and93}.

%
%
The number of atomic levels included for each atom/ion in
Table~\ref{tab:atomic_models} is fairly modest, although large enough
to include most of the collisionally-excited lines seen in the optical
and UV spectra of Be stars.  While non-LTE solutions can be sensitive
to the number of atomic levels included (see, for example, \cite{sig96}
for a case-study of C\,{\sc ii} in B~stars), the computational time also
increases sharply with the number of levels. Table~\ref{tab:atomic_models}
represents a compromise between realism and computational efficiency.
Several techniques exist to group atomic levels into ``super-levels"
\citep[see, for example,][]{hhl94} which will allow future work to
utilize more complete atomic models.

\begin{deluxetable}{rlrr}
\tablecaption{Atomic Models\label{tab:atomic_models}}
\tablehead{
\colhead{Z} & \colhead{Atom}    & \multicolumn{2}{c}{Number of} \\
\colhead{}  & \colhead{}        & \colhead{Levels} &
\colhead{Transitions}
}
\startdata
1 & H\,{\sc i}    & 15     & 77  \\
1 & H\,{\sc ii}   & 1      &  0  \\
2 & He\,{\sc i}   & 13     & 16  \\
2 & He\,{\sc ii}  & 1      &  0  \\
6 & C\,{\sc i}    & 15     & 23  \\
6 & C\,{\sc ii}   & 10     & 20  \\
6 & C\,{\sc iii}  & 21     & 45  \\
6 & C\,{\sc iv}   &  8     & 15  \\
6 & C\,{\sc v}    &  1     &  0  \\
7 & N\,{\sc i}    & 23     & 51  \\
7 & N\,{\sc ii}   & 15     & 24  \\
7 & N\,{\sc iii}  & 10     & 20  \\
7 & N\,{\sc iv}   &  1     &  0  \\
8 & O\,{\sc i}    & 15     & 19  \\
8 & O\,{\sc ii}   & 29     & 80  \\
8 & O\,{\sc iii}  & 15     & 24  \\
8 & O\,{\sc iv}   & 10     & 19  \\
8 & O\,{\sc v}    &  1     &  0  \\
12& Mg\,{\sc i}   & 17     & 28  \\
12& Mg\,{\sc ii}  & 12     & 25  \\
12& Mg\,{\sc iii} &  1     &  0  \\
14& Si\,{\sc i}   & 37     &109  \\
14& Si\,{\sc ii}  &  8     & 13  \\
14& Si\,{\sc iii} & 18     & 27  \\
14& Si\,{\sc iv}  & 11     & 23  \\
14& Si\,{\sc v}   &  1     &  0  \\
20& Ca\,{\sc i}   & 24     & 42  \\
20& Ca\,{\sc ii}  & 11     & 20  \\
20& Ca\,{\sc iii} &  1     &  0  \\
26& Fe\,{\sc i}   & 40     & 90  \\
26& Fe\,{\sc ii}  & 39     &191  \\
26& Fe\,{\sc iii} &  1     &  0  \\
  & {\sc total}   & 425    & 1001 \\
\enddata
\end{deluxetable}

\begin{deluxetable}{lr}
\tablewidth{0pt}
\tablecaption{Elemental Abundances\label{tab:abun}}
\tablehead{
\colhead{Element} & \colhead{Abundance\tablenotemark{a}}
}
\startdata
H  &  12.00 \\
He &  10.90 \\
C  &   8.55 \\
N  &   7.97 \\
O  &   8.87 \\
Mg &   7.58 \\
Si &   7.55 \\
Ca &   6.36 \\
Fe &   7.51 \\
\enddata
\tablenotetext{a}{The tabulated is abundance is $\log(N/N_{\rm H})+12$.}
\vspace{0.1in}
\end{deluxetable}

The density structure of the disk is chosen in an ad-hoc manner.
All calculations assume that the density
drops as an $R^{-n}$ power-law in the equatorial plane, and at each $R$, the
gas is in vertical,
isothermal hydrostatic equilibrium. To obtain the density 
above the
equatorial plane at each radial distance, the user supplies a fixed set of density-drops from
the equatorial plane, 
\begin{equation}
d_j\equiv\ln\{\rho(Z_j)/\rho(Z_j=0)\}
\end{equation}
for $j=1\ldots n_Z$. Here $d_1=0$ by definition, and $d_{n_z}\equiv-4$.  Then the $Z_j$
at which this density drop would occur, given the current value of $R_i$,
is computed assuming vertical hydrostatic equilibrium with an isothermal
temperature $T_o$,
\begin{equation}
\frac{Z_j}{R_i}=\left\{\left(\frac{\alpha /R_i}{d_j+
  \alpha /R_i}\right)^2-1\right\}^{\frac{1}{2}} \;.
\end{equation}
Here $\alpha$ is given by
\begin{equation}
\label{eq:scaleH}
\alpha=\frac{\mu\,m_{\rm H}}{k\,T_o}\,GM \,,
\end{equation}
where $\mu$ is the mean-molecular weight, $\approx 0.5$ for an ionized,
pure hydrogen disk,
and M is the mass of the
central B-star.  Thus the density at $(R_i,Z_j)$ is given by
\begin{equation}
\rho(R_i,Z_j)=\rho_0\left(\frac{R_i}{R_*}\right)^{-n}\,d_j \;.
\end{equation}
In this expression, $\rho_o$, $n$, and $T_o$ (through equation~\ref{eq:scaleH})
are the parameters which define the density structure of the disk.

At each computational grid point, the photoionizing
radiation field is required to evaluate the photoionization rates of
all atoms/ions for the statistical equilibrium equations and to
compute the photoionization heating rates for the radiative equilibrium
solution. It is usual to divide this radiation field into a direct and
diffuse contribution,
\begin{equation}
\label{eq:jnu}
J_{\nu}=J^{\rm Dir}_{\nu} + J^{\rm Dif}_{\nu} \;.
\end{equation}
The direct contribution represents the radiation from the central star
while the diffuse contribution arises from the disk. Note that despite
this division, the only energy input into the circumstellar disk is
assumed to be from the star itself so that ultimately, the energy in
the diffuse field has its origin in direct radiation from the star.

%
%
There is some evidence that $\gamma$~Cas has a binary companion with an
orbital period of $\sim200$ days \citep{har00,mir02}.  However, the exact
orbit and the nature of the companion, including its spectral type and
luminosity, are unknown. Rather than introduce further uncertain parameters
into the calculation (and invalidate the axisymmetric geometry), we
shall assume that the energy input into the disk of $\gamma$~Cas from
any potential companion is negligible.

The direct component from the central star to the photoionizing radiation
field at grid location $(R_i,Z_j)$ is given by
\begin{equation}
\label{eq:jdir}
J^{\rm Dir}_{\nu}(R_i,Z_j)=\int_{\Omega_*} I_{\mu\nu}(R_i,Z_j,\hat{n})\,d\Omega\,
\end{equation}
where $d\Omega$ is an infinitesimal patch of solid angle centred around
the direction $\hat{n}$. The integral is over the visible stellar surface.  
Typically the surface is divided into a few hundred patches and the transfer
equation is solved along a ray from the centre of each patch to the
grid location\footnote{By choosing to solve the transfer equation along these rays,
as opposed to simply applying exponential extinction of the stellar photospheric
intensity,
some contribution of the diffuse field is included.}. 
The radiation field at the stellar surface was taken
from an LTE stellar atmosphere of \citet{kur93} which specified
the mean intensity, $J_{\nu}(\tau_{\nu}=0)$, at the top of the photosphere over a grid
of 1221 frequencies.  This was turned into the required intensity at
each surface element by using the limb-darkening law
\begin{equation}
I_{\mu\nu}=I_o\,\left\{1-a_{\nu}(1-\mu)-b_{\nu}(1-\mu)^2\right\}
\end{equation}
where the coefficients
$a_{\nu}$ and $b_{\nu}$ were linearly interpolated from Table~V of
\citet{wad85}.  Here $\mu$ is the usual cosine of the surface viewing
angle. 
Computing the mean intensity from this expression
and setting it to the LTE model atmosphere prediction, we find that
\begin{equation}
I_o=\frac{J_{\nu}(\tau_{\nu}=0)}{1-a_{\nu}-(3/4)\,b_{\nu}} \,.
\end{equation}
While this procedure is approximate, and the exact $I_{\mu\nu}$
for each $\mu$ predicted by the LTE model could have been used, this approximation
seems commensurate with others made in the
construction of these models. We also note the use of a more physically
realistic non-LTE, line-blanketed atmosphere for $J_{\nu}(\tau_{\nu}=0)$
would be a useful future improvement.

%
%
In addition to limb darkening, gravity darkening induced by rapid stellar
rotation can change the intensity distribution across the stellar disk
and hence modify the direct component to the photoionizing radiation field.
\cite{tyc05} interferometrically resolved $\gamma$~Cas's disk and
estimated its inclination angle to the sky. They conclude the
$\gamma$~Cas rotates at $0.7\pm0.1$ of its critical velocity. As
$\gamma\,$Cas does not seem to rotate particularly close to its critical
velocity, we have not included gravity darkening (and the associated
geometrical distortion of it's surface) into the calculation of the
direct photoionizing radiation field. This point is further discussed
in Section~\ref{sec:MM} where the adopted stellar parameters for
$\gamma\,$Cas are discussed.

%
%
The simplest treatment for the diffuse field is to employ the
on-the-spot (OTS) approximation in which
the recombination rate to level $n$ of hydrogen is written as
\begin{equation}
\label{eq:OTS}
R_{\kappa,n}^{\rm H}(\tau_n)\,\equiv\,R_{\kappa,n}^{\rm H}\,e^{-\tau_n} \,.
\end{equation}
Here $\tau_n$ is the optical depth at the continuum limit for
photoionization from level $n$ along a vertical ray to the nearest edge
of the disk (this is consistent with our assumption that the dominant
photon escape route is perpendicular to the disk ---see later discussion).
Thus the principle assumption of the OTS approximation is that at
high continuum optical depths, $\tau_n\gg1$, recombination to level $n$
produces a photon that is locally absorbed within the same volume
element, essentially undoing the recombination. Including the continuum
optical depth dependence in equation~(\ref{eq:OTS}) ensures that the OTS
approximation is used only when $\tau_n\gg1$; the full recombination
coefficient is employed when the gas becomes optically thin in the
continuum. We have applied to OTS approximation only to recombination to
level $n=1$ in hydrogen.  The optical depths in the remaining continua
are typically not large enough for a significant effect.  We discuss
a more complex, but still approximate, treatment for the diffuse field
in section~\ref{sec:diffuse}.

Given the photoionizing radiation field and current estimates of the electron
temperature and electron density at each grid location, we solve the
statistical equilibrium equations to obtain the level populations for
all atoms and ions. For atom $k$, and all of its associated ions, these are
\begin{equation}
\sum_{j\ne i}^{N^k_L} n^k_i R_{ij} - \sum_{j\ne i}^{N^k_L} n^k_j R_{ji} = 0 \,,
\end{equation}
for $i=1,\ldots N^k_L$ where $N^k_L$ is the number of atomic levels included
for the $k^{\rm th}$ atom and its ionization stages.
Note that these equations must be supplemented by a particle
conservation equation of the form
\begin{equation}
\sum_{i}^{N^k_L} n^k_i = 10^{A_k-12} n_{\rm H}
\end{equation}
for each atomic species. The elemental
abundance, $A_k$, can be found
from Table~\ref{tab:abun}.

In the case of bound-bound transitions $i\rightarrow j$,
$i<j$, the rates have the simple form
\begin{equation}
R_{ij} = n_{\rm e}\,q_{ij}(T_e)
\end{equation}
and
\begin{equation}
R_{ji} = A_{ji}\,P_{\rm esc}(\tau_z) + n_{\rm e}\,q_{ji}(T_e) \,.
\end{equation}
Here the factors $q_{ij}(T_e)$ and $q_{ji}(T_e)$ represent collisional
excitation and de-excitation respectively and are proportional to
the Maxwellian-averaged collision strength for $i\,\rightarrow\,j$
transition.  $A_{ji}$ is the usual Einstein transition probability for
spontaneous emission. This form of the statistical equilibrium equations
handles radiative transfer in the line via the escape-probability
approximation. The escape probability for each grid location was obtained
by computing the line-centre optical depth to the nearest vertical edge
of the disk (denoted $\tau_z$) and then using this optical depth to
estimate the static, single-flight escape probability assuming complete
redistribution in the spectral line.

%
%
We have assumed that the dominant loss route for
photons is perpendicular to the disk because of the exponential density
stratification implied by vertical hydrostatic equilibrium. As it is
reasonable to assume that the main motion of the disk gas is Keplerian
rotation about the central star, the Doppler shifts experienced by photons
escaping roughly perpendicular to the disk will be small, and it is appropriate
to approximate the escape probability, $P_{\rm esc}(\tau_z)$, by a {\it
static\/}, single-flight  escape probability, as opposed to employing the
Sobolev approximation. This is consistent with the definition of $\tau_z$
as the optical along a ray perpendicular to the disk to the nearest edge;
$\tau_z$ is not defined in terms of a local velocity gradient as in the
Sobolev approximation.  
%

%
%
The form of the static, single-flight escape probability appropriate for
complete-redistribution over a Doppler profile is
\begin{equation}
P_{\rm esc}(\tau) = \frac{1} { 4\tau\left(\ln(\tau/\sqrt{\pi})\right)^{1/2} }\,,
\end{equation}
where $\tau\gg1$ \citep{mil78,can85}. As this result
is an asymptotic expansion \citep[essentially in the limit 
that the scale of variation
of the line source function is large compared to the width of the scattering
kernel -- see][]{can85}, an ad-hoc correction is needed to handle 
$\tau\ll1$. Here we have adopted the suggestion of \citet{tie05} where the
escape probability is taken to be
\begin{equation}
P_{\rm esc}(\tau) = \frac{1-e^{-2.34\tau}}{4.68\tau}\,,
\end{equation}
for $\tau<7$. This function is continuous with the asymptotic result at
$\tau=7$ and tends to the limit of $1/2$ as $\tau\rightarrow 0$, a reasonable
result if the collisional destruction probability of the line photon upon
scattering is not too small.

In the case $i\rightarrow j\equiv\kappa$ is a bound-free transition, the
photoionization and recombination (spontaneous plus stimulated) rates are given by
\begin{equation}
R_{i\,\kappa}=4\pi\int_{\nu_o}^{\infty}\,\sigma_{i\kappa}(\nu)\,J_{\nu}\, \frac{d\nu}{h\nu}
\end{equation}
and
\begin{equation}
\label{eq:recom}
R_{\kappa\,i}=4\pi\left(\frac{n_i}{n_{\kappa}}\right)^{*}
\int_{\nu_o}^{\infty} \sigma_{i\kappa}(\nu) \left(\frac{2h\nu^3}{c^2}+J_{\nu}\right)\,\frac{d\nu}{h\nu} \,.
\end{equation}
Here $(n_i/n_{\kappa})^*$ is the LTE population ratio found from the
Saha-Boltzmann equation and it is proportional to the electron density,
$n_e$ \citep{mil78}.  Dielectronic recombination and autoionization are
included by retaining the full resonance structure of the photoionization
cross section $\sigma_{i\kappa}(\nu)$.

Given the solution for the atomic level populations, a new estimate for
the electron density can be made by enforcing charge conservation, and
the rates of heating and cooling for the various atomic processes can
then be computed.  Heating includes photoionization and collisional
de-excitation while cooling includes radiative recombination and
collisional excitation. Detailed expressions for all of these processes
can be found in \cite{ost89}. However, in contrast to \citet{ost89},
transitions (and the implied cooling) due to radiative recombination
were computed explicitly via equation~(\ref{eq:recom}) for each atomic
level; total recombination co-efficients (summed over $n$) were not used.

Heating due to viscous dissipation in a Keplerian disk was also included
\citep{lee91} but was always found to be negligible.

Net cooling due to free-free emission (in the fields of H\,{\sc i}
and He\,{\sc ii}) was included via the expression of \citet{ryb79},
modified as suggested by \citet{net90} to account for the reduction in the
free-free cooling rate as the gas becomes optically thick to free-free
radiation,
\begin{equation}
L_{\rm ff}= 1.4\cdot10^{-27}\,\sqrt{T_e}\,n_e\, 
\sum_{i=\rm H, He}\,n_i\,Z_i^2\,\overline{g_{ff}}\,e^{-h\nu_{\rm max}/kT_e}\,.
\end{equation}
Here $Z_i=1$ and the frequency cut-off, $\nu_{\rm max}$, suggested by 
\citet{net90},
is the smallest frequency for which the optical depth to the nearest
edge of the disk exceeds one.  The Gaunt factor, $\overline{g_{ff}}$,
which varies slowly with temperature, was set to a constant value of
1.2 which \citet{ryb79} indicate will approximate the exact result to
within 20\%.

To find the equilibrium kinetic temperature, $T_e$, at each grid location,
heating and cooling were balanced by searching for a zero in
the net-cooling rate, $\eta_{\rm C}(T_e)$. The root was initially located via
bisection and then refined with the secant method. In the rare case of multiple
roots, the stable one satisfying $d\eta_{\rm C}/dT_e>0$ was chosen.

The overall flow of the calculation is to start at the inner boundary of
the disk, closest to the star at $i=1$. Solutions proceed downward in $Z$,
from the top of the disk ($j=n_z$) to the equatorial plane ($j=1)$. This
allows the optical depths back to the star and to the nearest edge of
the disk to be kept current with the solution level populations.

\section{Computations}

In this section, we first compare the predictions of our code with known
results for $\gamma\,$Cas. Next we explore a wide range of disk parameters
for $\gamma\,$Cas and investigate the effect on the temperature structure
of the disk of using a solar chemical composition for the gas. We then
examine the energy loss in the spectral lines included in the models
and compute the near infrared spectral energy distribution. Finally,
we examine the computation of the diffuse photoionizing radiation field
generated by the disk itself in order to evaluate the use of the OTS
approximation for most of the models computed in this work.

\subsection{Comparison with Millar \& Marlborough}
\label{sec:MM}

\citet{mil98} constructed a pure-hydrogen radiative equilibrium model
for $\gamma\,$~Cas. \citet{mil99} (MM, hereafter) extended this work
to include the OTS approximation for the diffuse radiation field,
and it is this temperature distribution which we have chosen for
comparison. We adopt the same stellar parameters for $\gamma\,$~Cas as
MM, which are reproduced in Table~\ref{tab:gamma_cas_star}. As we use
these fundamental parameters for all of the calculations in this work,
some additional comment is in order, particularly concerning the adopted
stellar effective temperature. \citet{fre05} investigate of the effect
of rapid rotation on fundamental parameter determinations of B~stars.
They find, by fitting the line spectrum between $4250$ and $4500\;$\AA,
``apparent stellar parameters" for $\gamma\,$Cas (those obtained by
a best fit classical, plane-parallel model atmosphere) of $T_{\rm
eff} = 26,400\;$K and $\log(g)=3.8$ which are close to the parameters
adopted in this work.  Nevertheless, they do find significant effects
of rotation in their best-fit rotating models which have a {\it parent
non-rotating counterpart\/} $T_{\rm eff}$ (see their paper for details)
of $\approx\,30,000\;$K. This result suggests that accounting for the
rotation of $\gamma\,$Cas in a manner following \citet{fre05} (but using
non-LTE stellar atmospheres) would be a useful future improvement in
the computation of the direct stellar contribution to the photoionizing
radiation field.

The fixed density structure for the $\gamma\,$~Cas disk adopted by MM
is described by \citet{mar69} and is slightly different from the model
described in section~\ref{sec:detail}. MM assume that the disk is in
isothermal, hydrostatic equilibrium at only one radial distance from
the central star and that the radial drop-off in the equatorial density
follows from an assumed (radial) outflow velocity law and the equation
of continuity.  We have simply adopted the $(R_i,Z_j)$ grid of MM (which
is 24 by 20) and their total density at each grid point as input to {\sc
bedisk}. We have used a 5-level hydrogen atom plus continuum for this
comparison and have also used the OTS approximation (as described previously)
for the diffuse field. Despite this, there are still some significant
differences between the calculations: our approach uses the optical depths
in the OTS approximation and in the line escape probabilities as opposed
to the various cases of MM. We use a newer ATLAS stellar atmosphere to
predict the photoionizing radiation field from the star and use many
more rays from each grid point back to the star. No attempt was made
to use identical atomic data: MM included collisional transitions for
only transitions $n$ to $n\pm 1$ in hydrogen whereas we have included
all collisional rates.  Nevertheless, despite these differences, the
comparison is a useful check.

\begin{deluxetable}{lrl}
\tablewidth{0pt}
\tablecaption{Stellar Parameters for $\gamma\,$ Cas.\label{tab:gamma_cas_star}}
\tablehead{
\colhead{Parameter} & \colhead{Value} & \colhead{Unit}
}
\startdata
Spectral Class& B0 IVe            & \nodata \\
Radius        & 10.0              & $R_{\sun}$\\
Mass          & 17.0              & $M_{\sun}$\\
Luminosity    & $3.4\,\,10^4$     & $L_{\sun}$\\
$T_{\rm eff}$ & 25,000            & K\\
$\log(g)$     &  3.50             & $\rm cm\,s^{-2}$~\\ 
Distance$^{a}$& $188^{+22}_{-18}$ & pc \\
\enddata
\vspace{0.1in}
\tablenotetext{a}{Distance from the Hipparcos catalogue \citep{per97}.}
\end{deluxetable}

{\sc bedisk} predicts a density-weighted average temperature,
defined as
\begin{equation}
\overline{T_\rho} \equiv \frac{1}{M} \int_{\rm Disk}\, T\,\rho\,dV \,,
\end{equation}
(where $M$ is the total mass of the disk), of $11\,300$~K. This is to be 
compared to the $14\,500$~K quoted by
MM (in their Table~1). This significant difference is simply one of
definition: the density-weighted average quoted by MM is actually defined
as $(\sum \rho_{ij}\,T_{ij})/\sum \rho_{ij}$ where the sum is over all
of the grid points in the calculation. Computing this quantity for the
current {\sc bedisk} model yields $13\,900$~K which agrees with the MM
result to within 5\%.

\begin{figure}
\epsscale{1.0}
\plotone{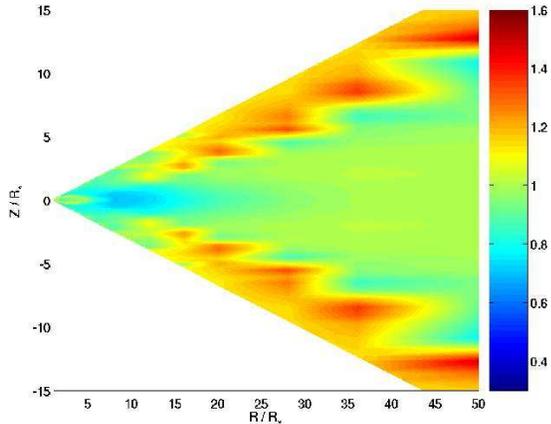}
\caption{Ratio of the {\sc bedisk} temperature to the MM temperature
for $\gamma\,$Cas.\label{fig:tratio_PM}}
\vspace{0.1in}
\end{figure}

Figure~\ref{fig:tratio_PM} compares the ratio of the {\sc bedisk}
temperature to the MM temperature throughout the entire circumstellar
disk.  Agreement is generally good; {\sc bedisk} tends to be somewhat
cooler near the equatorial plane in the inner portion of the disk, while
somewhat hotter towards the upper edge. However, 80\% of the grid points
agree to within $\pm20$\%.  The largest differences tend to occur along
with upper edge of the envelope where the optical depths (to the nearest
vertical edge of the disk) are most rapidly changing. The treatment of the
escape of line radiation, as noted above, and particularly of how the OTS
approximation was implemented (MM applied OTS to the whole disk as opposed
to including an optical depth dependence as in equation~\ref{eq:OTS}),
are likely the origin of the more significant differences.

\subsection{Effect of Adding Metals on the Thermal Structure}

Table~\ref{tab:models} gives the disk parameters for 16 disk
models computed to compare with observations of $\gamma$~Cas. 
These models span a range of nearly two
orders of magnitude in density (as obtained by varying $\rho_o$ from
$2.5\cdot\,10^{-12}$ to $1.0\cdot\,10^{-10}$ $\rm gm\,cm^{-3}$) with
two values assumed for the radial drop-off of the density in the equatorial plane,
$R^{-2.5}$ and $R^{-3.5}$. All models assumed
$T_o=13\,500\;$K for the isothermal temperature which sets the vertical
density scale-height via Eq.~\ref{eq:scaleH}. Also given in the table are the 
predicted density-weighted temperatures,
the disk emission measures (in $\rm cm^{-3}$), defined as
\begin{equation}
{\rm EM} \equiv \int_{\rm Disk} n_e^2 \,dV \,,
\end{equation}
and the predicted total H$\alpha$ luminosities (in $\rm ergs\,s^{-1}$).
Section~\ref{sec:lum} discusses how the H$\alpha$ luminosity was computed.
The parameters of the central star were again those of
Table~\ref{tab:gamma_cas_star}.

\begin{deluxetable}{ccccc}
\tablewidth{0pt}
\tablecaption{Model Disk Parameters.\label{tab:models}}
\tablehead{
\colhead{n} & $\rho_o$ & $\overline{T_\rho}$ & \colhead{$\log\,$EM} &
\colhead{$\log\,$H$\alpha$} \\
\colhead{}  & \colhead{$\rm g\,cm^{-3}$} & \colhead{K} & \colhead{$\rm cm^{-3}$} & 
\colhead{$\rm erg\,s^{-1}$}
}
\startdata
2.5 & $2.5\,10^{-12}$ & 14060 & 59.25 & 33.43 \\
2.5 & $5.0\,10^{-12}$ & 12870 & 59.84 & 33.79 \\
2.5 & $7.5\,10^{-12}$ & 12420 & 60.16 & 33.99 \\
2.5 & $1.0\,10^{-11}$ & 12170 & 60.40 & 34.12 \\
2.5 & $2.5\,10^{-11}$ & 11040 & 61.10 & 34.48 \\
2.5 & $5.0\,10^{-11}$ &  9420 & 61.50 & 34.69 \\
2.5 & $7.5\,10^{-11}$ &  8590 & 61.57 & 34.77 \\
2.5 & $1.0\,10^{-10}$ &  8140 & 61.62 & 34.80 \\
3.5 & $2.5\,10^{-12}$ & 13990 & 58.88 & 32.90 \\
3.5 & $5.0\,10^{-12}$ & 13740 & 59.48 & 33.17 \\
3.5 & $7.5\,10^{-12}$ & 13500 & 59.81 & 33.31 \\
3.5 & $1.0\,10^{-11}$ & 13290 & 60.05 & 33.40 \\
3.5 & $2.5\,10^{-11}$ & 12080 & 60.82 & 33.68 \\
3.5 & $5.0\,10^{-11}$ & 11050 & 61.35 & 33.85 \\
3.5 & $7.5\,10^{-11}$ & 10560 & 61.48 & 33.95 \\
3.5 & $1.0\,10^{-10}$ & 10240 & 61.55 & 34.01 \\
\enddata
\end{deluxetable}

\begin{figure}
\epsscale{1.0}
\plotone{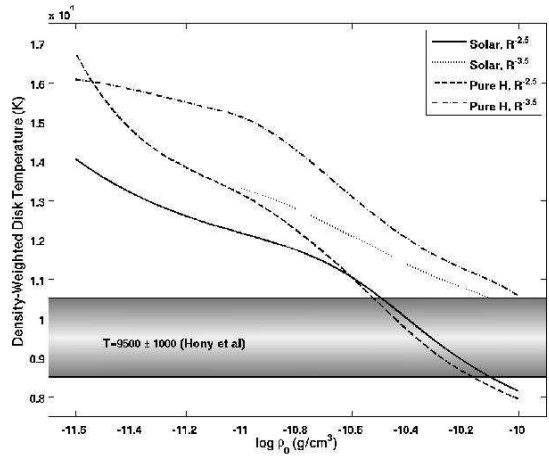}
\caption{Density-weighted disk temperatures for the models
of Table~\ref{tab:models}.\label{fig:denaverage}}
\vspace{0.1in}
\end{figure}

The density-weighted temperatures predicted by the models are plotted
in Figure~\ref{fig:denaverage} as a function of $\rho_o$. These results
are compared to the predicted density-weighted temperatures for a set
of pure hydrogen disks with identical physical parameters.  Also shown
in the Figure is the observed disk temperature for $\gamma$~Cas of
$9500\pm1000$~K as found by \citet{hon00} by fitting the IR Humphrey's
bound-free jump at 3.4$\,\mu$m. As expected, denser disks predict lower
density-weighted temperatures.

The $R^{-2.5}$ models are consistent with the \citet{hon00} result for
densities in the range of $3$ to $8\cdot\,10^{-11}\,\rm g\,cm^{-3}$. This
agrees well with the density estimated by \citet{hon00} using their
observations and the disk models of \citet{wat86}. The $R^{-3.5}$
models are only just consistent with the \citet{hon00} result for largest
densities considered, $\rho_o\approx 10^{-11}\,\rm g\,cm^{-3}$

The predicted temperature trend as a function of $\rho_o$ has an
interesting dependence on the metallicity of the gas. For low density
disks, the solar composition disks are considerably cooler than the
pure hydrogen models by 1-2000~K. However, the difference decreases for
higher disk densities, $\rho_o>2\cdot\,10^{-11}\rm \,g\,cm^{3}$.  Indeed,
for the $R^{-2.5}$ models, the higher density solar and pure hydrogen
disks predict nearly the same density-weighted temperatures. This
behaviour can be understood in terms of the heating and cooling avenues
introduced by metals. Metals can act to cool the gas due to the escape
of collisionally-excited line radiation. However, metals can also
help to heat the gas via photoionization. If the optical depths in
the hydrogen continua (excluding the Lyman continuum) are low, then
the additional heating provided by the photoionization of metals is
negligible in comparison to hydrogen. In this case, it is the cooling due
to collisionally-excited line radiation that dominates. In a low-density
disk, line cooling is further enhanced by the small optical depths
(and hence high escape-probabilities) in the lines.

At high densities, however, the optical depths in the hydrogen bound-free
continua are much larger; photoionization heating due to metals can then
become important, particularly as many abundant metals have bound-free
thresholds in the short-wavelength region of the Balmer continuum which
is near the photospheric flux maximum in B~stars. In this case, metals
add both heating and cooling and the net result is a very similar
density-weighted temperature to the case of a pure hydrogen plasma.
These trends help explain why \citep{mil98} where able to obtain a
reasonable density-weighted temperature for $\gamma$~Cas despite using
a pure hydrogen envelope.

\begin{figure}
\epsscale{1.0}
\plotone{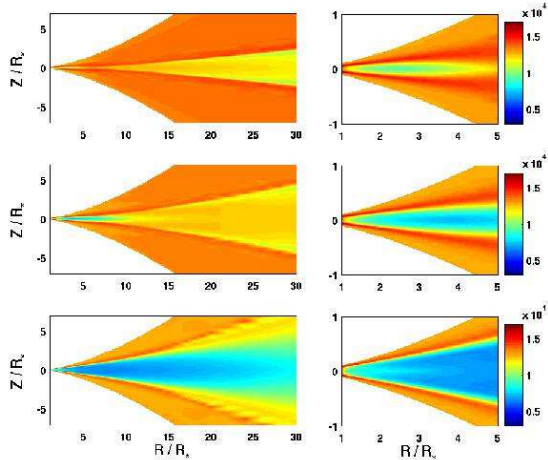}
\caption{Circumstellar disk temperature as predicted by three $R^{-2.5}$
models. The top panel has $\rho_o=1\cdot\,10^{-12}\,\rm g\,cm^{-3}$,
the middle panel, $\rho_o=5\cdot\,10^{-12}$, and the bottom panel,
$\rho_o=5\cdot\,10^{-11}$. Each panel on the right enlarges the inner
portion of the disk from the panel on its left.
\label{fig:2Dtemp}}
\vspace{0.1in}
\end{figure}

Two-dimensional temperature distributions in the disk for
several different density $R^{-2.5}$ models are shown in
Figure~\ref{fig:2Dtemp}. The inner portion of the disk for $R/R_*<5$
is expanded in each case for clarity. These figures clearly show the
development of a cool region near and in the equatorial plane for the
higher density disks.  These denser disks have fairly strong vertical
temperature gradients perpendicular to the equatorial plane. Given
this, it would seem prudent to re-integrate the equation of hydrostatic
equilibrium at each radial distance, accounting for the vertical variation
of the gas temperature, and then to iterate the pressure structure along
with the thermal solution to produce a disk that is in both radiative and
(vertical) hydrostatic equilibrium; we shall present such models in a
future work \citep{mcg06}. However, as the focus of the present work is
on the thermal structure of the disk, it is convenient to have a fixed
density structure so that the thermal effect of the gas metallicity can
be unambiguously seen.

\begin{figure}
\epsscale{1.0}
\plotone{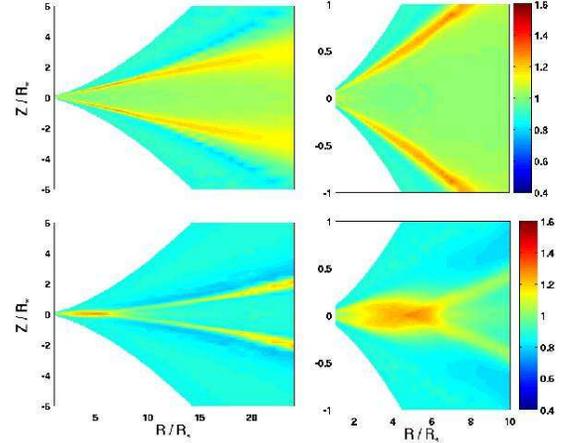}
\caption{Temperature ratio between a full solar composition disk to that
of a pure hydrogen disk. Shown in the bottom panel is a model
with $\rho_o=5\cdot\,10^{-12}\,\rm g\,cm^{-3}$, and in the top panel,
one with $\rho_o=5\cdot\,10^{-11}\,\rm g\,cm^{-3}$.
Both models had an $R^{-2.5}$ density drop-off in the equatorial
plane. Each panel on the right enlarges the inner
portion of the disk from the panel on its left.
\label{fig:add_metals}}
\vspace{0.1in}
\end{figure}

Figure~\ref{fig:add_metals} presents a detailed comparison of the
temperature structure predicted by a gas with a realistic
solar composition to that of a pure hydrogen model.  The figure
compares the temperature ratio $T^{\rm Solar}/T^{\rm Pure H}$
for two densities, $\rho_o=5\cdot\,10^{-11}$ and $\rho_o=5\cdot\,10^{-12}\,\rm
g\,cm^{-3}$. In both cases, the optically thin gas far above the
equatorial plane is cooler with the inclusion of metals. However, in the
lower density model, $\rho_o=5\cdot\,10^{-12}\,\rm g\,cm^{-3}$, the gas in
the equatorial plane near the star is hotter for the solar composition
gas out to $R/R_*\sim 6$. In the higher density model, the gas near the
equatorial plane has nearly the same temperature in the solar and pure
hydrogen models. These trends are consistent with the effects noted
in the discussion of the density-weighted average disk temperatures.

\subsection{Line Luminosities}

\label{sec:lum}

\begin{figure}
\epsscale{1.0}
\plotone{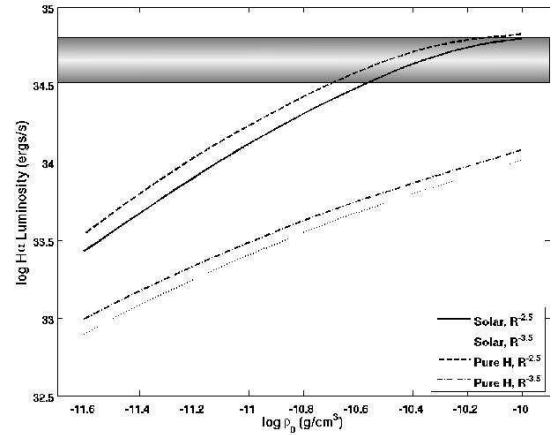}
\caption{The predicted energy loss in H$\alpha$ from
Eq.~\protect\ref{eq:lum} (in $\rm ergs\,s^{-1}$)
for the models of Table~\protect\ref{tab:models}. Shown as the shaded rectangle
is the observed range of values from \cite{kas89} and \citep{ste95}.
\label{fig:halpha_flux}}
\vspace{0.1in}
\end{figure}

Figure~\ref{fig:halpha_flux} plots the total energy lost in
H$\alpha$ (in $\rm ergs\,s^{-1}$) by the 16 models of Table~\ref{tab:models}.
The line luminosity was found by integrating the flux divergence, in the 
escape probability approximation\footnote{In the escape probability
approximation, the net radiative bracket is replaced by the (single-flight)
escape probability.}, 
over the volume of the disk, {\it i.e.}
\begin{equation}
\label{eq:lum}
L=h\nu_{ij} A_{ji}\int_{\rm Disk} n_j\,P_{\rm esc}(\tau_{ij}) \, dV \,.
\end{equation}
For H$\alpha$, $i$ is level $n=2$ of H\,{\sc i} and $j$ is level $n=3$.
Also shown in the Figure is the H$\alpha$ luminosity observationally determined
by \citet{kas89} and \citet{ste95}. The H$\alpha$ luminosity from $\gamma\,$Cas
is known to be variable, but the cited values are typical of the current
epoch. There is good agreement with the observed luminosity for
$R^{-2.5}$ models with $\rho_o$ between $2.5\cdot\,10^{-11}$ 
and $10^{-11}\,\rm g\,cm^{-3}$. This result is
consistent with the models that best fit the observed disk temperature
of \cite{hon00}. However, the $R^{-3.5}$ models seem inconsistent with the
total energy loss in H$\alpha$ for the range of disk densities considered.

\begin{figure}
\epsscale{1.0}
\plotone{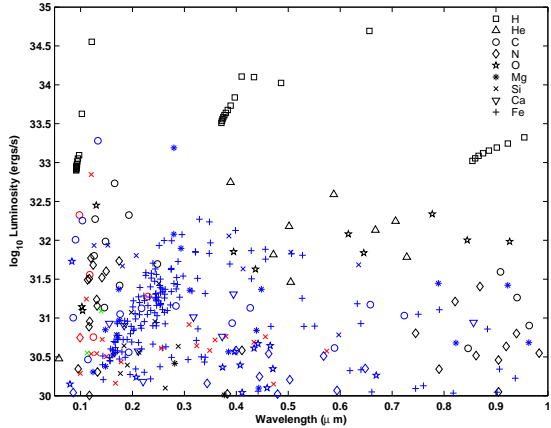}
\caption{The energy loss in each radiative transition predicted
by the $R^{-2.5}$ disk model with $\rho_o=5\cdot\,10^{-11}\rm\,g\,cm^{-3}$
Only transitions with an energy loss above $10^{30}\rm\,ergs\,s^{-1}$
are shown. The symbols representing the various elements are as
indicated in the legend. The ionization stage
is coded by colour: neutral (black), first ion (blue), second ion (red),
third ion (green).
\label{fig:line_flux}}
\vspace{0.1in}
\end{figure}

Figure~\ref{fig:line_flux} shows energy escaping in all of the
included radiative transitions for the $R^{-2.5}$ model with
$\rho_o=5\cdot\,10^{-11}\,\rm g\,cm^{-3}$. The fluxes are again found by
integrating Eq.~\ref{eq:lum} over the disk.  It is important to keep in
mind that Figure~\ref{fig:line_flux} is not a spectrum (which would be
obtained by integrating the transfer equation along a series of rays
through the computational domain); it is simply a plot of the energy
loss per second in each line acting to cool the gas. Nevertheless,
it gives a good indication of the expected strong emission lines in
the disk spectrum. For this particular model, 94\% of the energy loss
is provided by the lines of H\,{\sc i}. Contributing at the level of
$\approx\,1$\% percent are Fe\,{\sc ii}, C\,{\sc ii}, Mg\,{\sc ii}, and
He\,{\sc i}. Next to the lines of H\,{\sc i}, the largest energy losses
are in the resonance lines $\lambda\,1333.6\,$\AA\ line of C\,{\sc ii}
($\rm 2s^2\,2p\,^2P^o\,-\,2s\,2p^2\,^2D$) and the h and k lines of
Mg\,{\sc ii} near $\lambda\,2800\,$\AA. Although it does not possess a
single strong line, cooling due to Fe\,{\sc ii} dominates over all of
the metals due to its rich spectrum with many collisionally-excited lines.
It should be noted that Figure~\ref{fig:line_flux} and the percentage
contributons cited above represent a {\it global\/}
picture of energy loss integrated over the entire disk. The impact of the
heating and cooling contributions of metals can be much larger at
individual grid locations as demonstrated by Figure~\ref{fig:add_metals}.

\subsection{Predicted IR spectral energy distributions}

In this section, we consider the predicted IR continuum energy
distribution as emitted {\it perpendicular\/} to the disk\footnote{We
shall present detailed synthesis for the infrared hydrogen disk spectrum for arbitrary
inclination angles in a future work.}. Such models have
a zero inclination angle between the rotation axis and the observer's
line of sight. This spectrum is easily found by solving the radiative
transfer equation vertically through the disk at each $R_i$. If
$I_i\equiv I_{+1,\nu}(R_i)$ is the emergent
intensity for the annulus of area $A_i=\pi(R^2_{i+1/2}-R^2_{i-1/2})$
then the SED seen by the observer will be
\begin{equation}
L^{\rm Star+Disk}_{\nu}=I^{*}_{\nu}\,\pi R^{2}_{*} + \sum_{i=1}^{n_{R}}\,I_i A_i\,.
\end{equation}
Here $R_*$ is the radius of the star, $I^*_{\nu}$ is the specific
intensity corresponding to the stellar surface, and $I_i$ is the specific
intensity of the $i^{th}$ disk annulus. It is well known that at IR
wavelengths, Be stars possess an excess of radiation over that predicted
by an appropriate stellar photosphere model \citep{cot87}.  This excess
comes mainly from free-free emission in the ionized disk\footnote{Note
that the exact expression for the free-free Gaunt factor was used for
this calculation.}, with a small contribution from free-bound emission
shortward (in wavelength) of the ionization edges.

\begin{figure}
\epsscale{1.0}
\plotone{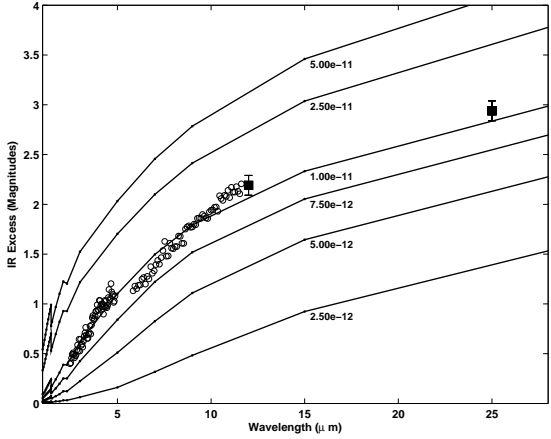}
\caption{The IR excess expressed in magnitudes predicted by the
$R^{-2.5}$ models of Table~\protect\ref{tab:models}. The square data points at
$12$ and $25\,\mu$m are from \protect\citet{cot87}. The open circles are the
ISO spectrophotometry.
\label{fig:IR}}
\vspace{0.1in}
\end{figure}

The near-IR spectral energy distributions are shown in Figure~\ref{fig:IR}
for the $R^{-2.5}$ models of Table~\ref{tab:models}.  Plotted is the
(monochromatic) IR excess expressed in magnitudes,
\begin{equation}
Z_{\nu}\equiv 2.5 \log\left(\frac{L^{Star+Disk}_{\nu}}{L^{Star}_{\nu}}\right).
\end{equation}
The discontinuous jumps at wavelengths less than ~$5\,\mu$m in this
figure, particularly for higher densities, represent the hydrogen
free-bound continua for recombination to $n=4$ ($1.4\,\mu$m) and $n=5$
($2.2\,\mu$m). Unlike (plane-parallel) stellar photospheres which
have bound-free edges in absorption (reflecting the inward increase
in temperature), circumstellar material exhibits the bound-free edges
in emission, reflecting the increase in the gas emissivity due to
recombination. Thus the figure indicates a small contribution from
free-bound emission to the infrared excess at these wavelengths. We also
note that as the disk density is increased, the infrared excess steepens
most strongly between the wavelengths of 1 and 5$\;\mu$m.

To compare with observations, we first show the $12$ and $25\,\mu$m IRAS
IR excesses for $\gamma\,$Cas found by \citet{cot87}.  The model that
best matches the IRAS IR excess at these wavelengths has a slightly
smaller density, $\rho_o\approx 10^{-11}\,\rm g\,cm^{-3}$, than the
model that best matches the observed average disk temperature and energy
loss in H$\alpha$, $\rho_o\approx 3\cdot\,10^{-11}\,\rm g\,cm^{-3}$
(Figure~\ref{fig:denaverage}).  However, there is good evidence
that $\gamma\,$Cas is seen at an inclination of $i>\approx\,55^o$
\citep{tyc06}, as opposed to being viewed pole-on ($i=0^o$) as assumed
by the models.  Hence the models have an effective emitting area that is
too large and the IR excess is likely overestimated.  In addition, the
spectra for non-zero inclination angles will reflect the contribution of
a different set of rays passing through the disk; the different physical
conditions along these rays will lead to differences in the gas opacity
and emissivity and hence a different predicted intensity along each ray.

We also show in Figure~\ref{fig:IR} the 2.5 to 11.6$\,\mu$m Infrared Space
Observatory (ISO) spectrophotometry\footnote{We have extracted this data
from the ISO on-line archive. The spectrophotometry is from the PHT-40
instrument for observing series TDT~76803401.} for $\gamma\,$Cas. In
order to compare the ISO fluxes at the Earth to our photospheric model
(to set the reference flux to extract the excess), we require the angular
diameter of the star.  \citet{tyc05} cite the major axis of the H$\alpha$
emitting region of $\gamma\,$Cas of $3.67\pm0.09\,$milli-arcseconds,
but this has a large contribution from the circumstellar disk. To
resolve this problem, we have chosen an angular diameter such
that the ISO data reproduces the $12\,\mu$m IRAS excess. Combining
this stellar angular diameter for $\gamma\,$Cas with the range of
Hipparcos distances listed in Table~\ref{tab:gamma_cas_star}, we find a
required radius for $\gamma\,$Cas of between 6.8 and 8.4 solar radii.
While this result in disagreement with the 10 solar radii listed in
Table~\ref{tab:gamma_cas_star}, the larger value is within the plausible
error range for $\gamma\,$Cas's radius.  With this normalization,
the ISO observations match reasonably well with the $\rho_o\approx
10^{-11}\,\rm g\,cm^{-3}$ model prediction over the considered wavelength
range. The fit is actually worse closer to the normalization point; the
shorter wavelength excess, between $2.5<\lambda<5\,\mu$m, is quite well
reproduced. Similar caveats to those ending the previous paragraph apply
here: a detailed comparison will be made with a model that correctly
accounts for the viewing inclination of $\gamma\,$Cas in a future work.

\subsection{An Approximate Treatment of the Diffuse Radiation Field}
\label{sec:diffuse}

While the OTS approximation is simplest treatment of the diffuse
radiation field, Figure~\ref{fig:2Dtemp} shows that in the densest disks
($\rho_o\approx 5\cdot\,10^{-11}\rm g\,cm^{-3}$), grid locations near
the equatorial plane can become quite cool due to large optical depths
back to the central star. However, examining the thermal structure
in the Z-direction at such a location typically shows that at heights
above and below the equatorial plane, the gas can still be quite hot
as it can be directly illuminated by at least a portion of the star.
One might expect that the radiation emitted from these hot ``sheaths"
might be an important source of secondary photoionizing radiation for
the equatorial gas \citep{car05}.

In principle,
the calculation of $J^{\rm Dif}_{\nu}$ in Eq.~\ref{eq:jnu} requires
a solution of the transfer equation in the 2D cylindrical
geometry\footnote{A direct implementation of this approach can be found in \citet{car05}
who use a Monte Carlo approach to radiative transfer to estimate the
diffuse radiation field in the envelope of a somewhat later-type Be star.}.
However, to estimate the potential effect of this diffuse field on 
the current work, we have tried a simpler and approximate approach.
We solve, at each $R_i$, the radiative transfer equation perpendicular to
the disk in the Z-direction using the method of short-characteristics
\citep{ols87}.  The mean intensity along this ray is then
\begin{equation}
J_{\nu}(Z_j)=\frac{1}{2}\left\{I_{\mu=+1,\nu}(Z_j)+I_{\mu=-1,\nu}(Z_j)\right\} \,.
\end{equation}
For this solution, zero incident radiation is assumed perpendicular to
the disk, We then estimate the diffuse contribution to the photoionizing
radiation field at each $Z_j$ by assigning
\begin{equation}
\label{eq:add_diff}
J^{\rm Dif}_{\nu}(Z_j)=W_{\rm dif}\,J_{\nu}(Z_j) \,,
\end{equation}
Here $W_{\rm dif}$ represents an ad-hoc dilution factor, between zero
and one, for the perpendicular rays. The expectation is that $J^{\rm Dif}_{\nu}$
will potentially be most important in the cool equatorial regions that
develop for higher density disks such as the one illustrated in
the bottom panel of Figure~\ref{fig:2Dtemp}. In such a cool obscured region,
the hot sheaths above and below cover approximately $1/2$ of the available $4\pi$
steradians so a dilution factor of $W_{\rm dif}=0.5$ is a reasonable approximation.
In regions of the disk where the central star is not obscured, the choice of
$W_{\rm dif}$ is irrelevant as $J^{\rm Dir}\gg J^{\rm Dif}$. For this reason,
we have taken $W_{\rm dif}$ to be $1/2$ at all grid locations in the disk. 

\begin{figure}
\epsscale{1.0}
\plotone{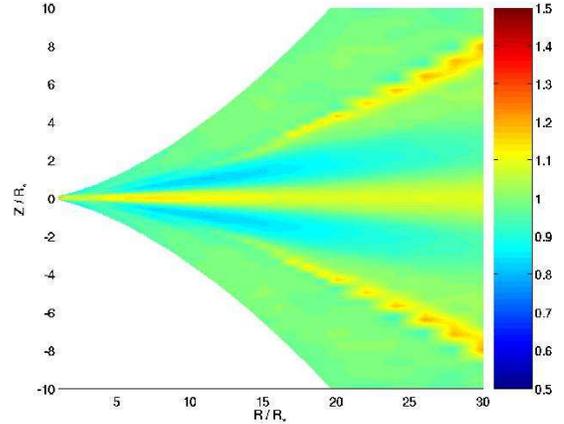}
\caption{The effect of adding the diffuse contribution of Eq.~\ref{eq:add_diff} 
to the temperatures predicted for
the $R^{-2.5}$ disk model with $\rho_o=5\cdot\,10^{-11}\rm\,g\,cm^{-3}$. Plotted is
the temperature ratio of a calculation including Eq.~\ref{eq:add_diff}
to one omitting this contribution.
\label{fig:add_diff}}
\vspace{0.1in}
\end{figure}

Figure~\ref{fig:add_diff} shows the effect of adding such an
approximate treatment of $J^{\rm Dif}$ to the $R^{-2.5}$ disk model with
$\rho_o=5\cdot\,10^{-11}\rm\,g\,cm^{-3}$, the model which best matched the
observed disk temperature and the total H$\alpha$ luminosity. With the
diffuse field added, the amount of additional heating in the equatorial
plane is fairly small, with only a $10-20$\% increase in the temperature
there.  Hence, we feel that the previously computed models are reliable
despite the approximate treatment of the diffuse field through the OTS
approximation.

Note that this figure is not meant to imply that the effect of the
diffuse field is negligible; in both runs, the on-the-stop (OTS)
approximation, which approximates the local trapping of radiation with
$\lambda<912\,$\AA\ and hence the diffuse photoionizing radiation field
for $\lambda<912\,$\AA\ was employed following Eq.~\ref{eq:OTS} If the
OTS approximation is turned off, there is considerable cooling of the
equatorial regions\footnote{A subtle point is whether a model including
$J^{\rm Dif}$ according to Eq.~\ref{eq:add_diff} should also employ
the OTS approximation.  In principle, employing both over counts the
local radiation field in each volume element.} \citep[see also][]{mil99}
and this would mask the effect (illumination of these same regions from
above and below) that we were trying to investigate.

\section{Conclusions}

For the first time, we have constructed radiative equilibrium models for
the disk of $\gamma\,$Cas using a gas with a solar chemical
composition. We find that a model with a power-law equatorial gas
density of $\rho(R)\approx 3$ to $5\cdot 10^{-11} (R/R_*)^{-2.5}\, \rm
g\,cm^{-3}$, which is in vertical (isothermal) hydrostatic equilibrium,
can reproduce several overall observed disk properties, including its
(density-weighted) average temperature, the energy loss in H$\alpha$,
and the near infrared flux excess.

Also in this work, we have investigated the differences between our solar
composition radiative equilibrium models and the corresponding set of
pure hydrogen models with the same disk parameters.  The effect of the
additional heating and cooling provided by elements heavier than hydrogen
on global measures of the disk structure, such as the density-weighted
average disk temperature or the total energy loss in H$\alpha$, depend
strongly on the parameter $\rho_o$ (the assumed equatorial density near
the stellar surface) which sets the overall density of the disk.  For low
disk densities, $\rho_o< ~ 10^{-11}\rm\,g\,cm^{-3}$, the density-weighted
averaged disk temperature is generally 1--2000~K cooler with a solar
composition, due mostly to enhanced collisionally-excited line cooling. At
higher densities, absorption of ionizing radiation in the bound-free
continua of elements heavier than hydrogen provides additional heating
which offsets somewhat the additional line cooling.

In examining the detailed temperature at each position withing the disk,
differences of up to $\approx\pm\,40$\% are found in comparing the solar
composition models to the pure hydrogen models.  While the global energy
loss in H$\alpha$ in the solar and pure hydrogen models is similar, these
significant differences in the disk temperature distribution may manifest
themselves in the detailed spectra.  The next step in the comparison
between these two sets of models is to present predicted spectra over
a wide range of wavelengths for a wide range of viewing inclinations.

The {\sc bedisk} code represents a compromise between computational
efficiency and realism. The code's most notable approximations are the
treatment of the disk diffuse radiation field, the use of (first-order)
static, escape probabilities for the line radiative transfer, the
assumption of a spherical, non-rotating, central star, and the somewhat
limited atomic models. Nevertheless, these approximations allow the code
to efficiently explore a wide region of parameter space for the thermal
structure of the circumstellar disks surrounding hot stars.  Future work
will proceed along two fronts: we will use the current version of {\sc
bedisk} to compute a series of Be disk thermal models that will serve
as the basis for detailed spectral synthesis to produce line profiles,
interferometric visibilities, and continuum polarization signatures which
can be directly compared to observations. Along the second front, we shall
improve some of the basic assumptions of the code. Most notably, we will
enforce a vertical hydrostatic equilibrium density structure consistent
with the thermal solution, allow for potential gravitational darkening
(and geometric distortion) of the central star by rapid rotation, and
account for the diffuse photoionizing radiation field from the disk by
a direct solution of the 2-D radiative transfer problem.

\acknowledgments

We would like to thank Chris Tycner for many helpful discussions. We
thank Rens Waters for clarifying the IR excess of $\gamma\,$Cas. We
also thank Kyle Lawson, Anna Molak, and Laura Thomson for help as part
of their NSERC Summer Student awards. Finally, we would like to thank
the anonymous referee for a careful reading of the text and providing
many helpful comments and suggestions.  This work is supported by the
Canadian Natural Sciences and Engineering Research Council through
Discovery Grants to TAAS and CEJ.

\appendix

\section{The Atomic Data}

Energy levels for all atoms and ions
were adopted from the NIST Atomic Spectra
Database\footnote{http://units.nist.gov/PhysRefData/ASD/index.html}.
Radiative transition probabilities and photoionization cross
sections were adopted from the Opacity Project Database TOPbase
\footnote{http://cdsweb.u-strasbg.fr/topbase.html}.  The photoionization
cross sections, which include complex resonance features due to
autoionization, were smoothed by convolution with a Gaussian down to the
resolution of the ATLAS frequency grid.  Thermally-averaged collision
strengths for the electron impact excitation of hydrogen were adopted
from \citet{cal94}, \citet{agg91}, and \citet{cha91}.  Thermally-averaged
collisional strengths for the remaining atoms and ions were adopted
from the compilation of \citet{pp95} where available. The Gaunt factor
approximation was used for the remaining dipole-allowed transitions. The
atomic data for iron and its ions were adopted from the extensive model
atoms constructed by \citet{sig03} and \citet{sig04}.

\end{document}